\documentclass[12pt]{article}
\def\ave#1{\left<{#1}\right>}
\usepackage{epsf,pstricks}
\input epsf
\newcommand{\forget}[1]{}
\begin{document}

\title{Transmission and Reflection in a Double Potential Well: Doing it the Bohmian
Way} \author{Regien G. Stomphorst \thanks{ E-mail:
regien@stomphorst.net}\\
 Molecular Physics Group, Dept. of Biomolecular Sciences, \\
Wageningen University,\\
Dreijenlaan 3,
6703 HA Wageningen
The Netherlands}
\maketitle

\begin{abstract}

The Bohm interpretation of quantum
mechanics is
applied to a transmission and reflection process in a double potential well. We
consider a time dependent periodic wave function
and study
the particle trajectories.
 The
average time, eventally transmitted particles stay inside the
barrier is the average transmission time, which can be
defined using the causal
interpretation. The
question remains whether these
transmission times
can be experimentally measured.
\end{abstract}

\section*{keywords}
Bohm interpretation; transmission times. 03.65 BZ

 \section{Introduction}

It is well known that (under certain circumstances) the causal
interpretation is empirically
equivalent
to the orthodox Copenhagen interpretation \cite{Cushing:96a,
Leavens:96a}.
However, there are a number of physical problems for which this
orthodox
approach provides no clear-cut answers. One of such problems is
the question of a time observable for a tunneling process.\\

Tunneling is the quantum mechanical phenomenon that a particle
can cross
a barrier with potential $V$, even if its energy is strictly
less
than
$V$. It is a natural question to ask how long it takes on
average
for
particles to cross such a barrier. Unfortunately
quantum
theory does not provide a suitable time operator, whose
expectation value
for a
given wave packet can be compared with experiment. Time
enters
quantum mechanics as a parameter, not as an operator. Therefore,
this
question about tunneling time is not an easy one.
\\

Tunneling processes
may be classified in two types: scattering type, where a wave
packet is
initially incident on a barrier, and then partly transmitted;
and
decay
type when, the particle is initially in a bound state,
surrounded
by a
barrier, and subsequently leaks out of this confinement.  Many
authors
have addressed the duration of tunneling
processes, in
case of scattering processes \cite{Hauge:89, Olkhovsky:92,
Landauer:94} and in case of a decay process \cite{Nogami:00}.\\

In Bohm's causal interpretation of quantum mechanics various
concepts of tunneling times for scattering processes can be
distinguished. The most well-known time is dwell time, i.e. the
time particles spend inside the barrier. For an ensemble of
particles, we can determine the average dwell time. This average
dwell time ($\ave{t_d}$) can be decomposed into an average
transmission times ($\ave{t_t}$), i.e. the time spent inside the
barrier by those particles, which eventually cross the barrier,
and average reflection time ($\ave{t_r}$), i.e. the time consumed
by reflected particles \cite{Leavens:96a}:
\begin{eqnarray}
\label{eq2.1}
\ave{t_d}=\left| T \right|^2 \ave{t_t}+\left| R
\right|^2
\ave{t_r} \end{eqnarray}
where $\left| T \right|^2$ and $\left| R \right|^2$ are
the transmission and reflection probabilities respectively. 
This relation consists of:
\begin{eqnarray}
\label{eq2.1a}
\ave{t_d}=\left| T \right|^2 \ave{t_t}+\left| R^{'}
\right|^2
\ave{t_{r^{'}}} +\left| R^{''}
\right|^2
\ave{t_{r^{''}}} \end{eqnarray}

where $R^{'}$ and $t_{r^{'}}$ refer to particles, which penetrate the barrier but re-emerge on
the same side and $R^{''}$ and $t_r^{''}$ to particles that do not enter 
the barrier.
This
relation
assumes that particles are either transmitted or reflected, i.e.
they
do not remain inside the barrier.
\\

 In this paper, we consider a decay type of transmission and 
reflection by applying
the causal interpretation to a double potential well. The
decomposition of the dwell time into average transmission and
reflection time according to relation (\ref{eq2.1}) in a decay
type of tunnelling is not as straightforward as for the scattering
cases. Definitions for transmission and reflection probabilities
are not common for this decay type of transmission and reflection.
 However, we shall
propose natural definitions for these concepts and use these to
define average transmission and reflection time. For reflection we will 
concentrate on
particles, which penetrate the barrier but re-emerge on
the same side. It will again
appear to be convenient to obtain the  average transmission time
by using the concept of the average arrival time.\\

The aim of this paper is firstly, to investigate whether the
causal interpretation of quantum mechanics provides a
straightforward way to define transmission times in double
potential wells.  For simplicity, we will study periodic wave
functions. Secondly, the question whether it is
necessary to adopt the causal interpretation to give meaning to
the thus obtained transmission times is addressed. The paper is
organised as follows.
 In section
\ref{sec2.3}
we apply
the causal interpretation method to transmission and reflection in a double
potential
well. In section \ref{sec1.5} we define transmission times in
terms of the probability density of the wave function.
 The last section is devoted to a discussion about the question
whether or not the causal interpretation provides a clear way to
define transmission times in a double potential well and about
the necessity to rely on the causal interpretation for this
definition.

\section{\label{sec2.3} The causal interpretation applied to a
double potential well}
 In this section, the causal interpretation is applied to
transmission and reflection in a double potential well. We discuss the possibilities
to define transmission and reflection coefficients, average dwell
time, average transmission time and reflection time according to
this method.

\subsection{Description of the double potential well}

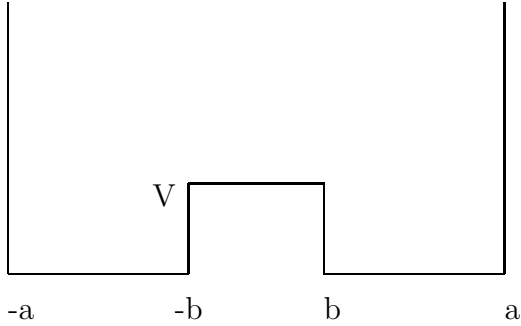
\begin{figure}
\setlength{\unitlength}{0.6cm}
\begin{picture}(11,7)
\put(0,0){-a}
\put(0,1){\line(0,1){6}}
\put(0,1){\line(1,0){4}}
\put(4,1){\line(0,1){2}}
\put(4,3){\line(1,0){3}}
\put(7,1){\line(0,1){2}}
\put(7,1){\line(1,0){4}}
\put(11,1){\line(0,1){6}}
\put(11,0){a}
\put(3.7,0){-b}
\put(7,0){b}
\put(3.2,2.5){V}
\end{picture}
\caption{\label{fig2.1} The system under consideration: a double
potential
well. The total length of the box is 2$a$, the potential at $a$
and $-a$ is infinite. The barrier is situated from$-b$ to $b$ and
has a constant height V.}
\end{figure}
To model
the decay type of transmission processes we consider a wave packet
in a double potential well.
 The potential well is described by a
one-dimensional box, defined from $-a$ to $a$ (see
Fig.~\ref{fig2.1}). At these points, the walls are infinitely
high. In the middle of the box a barrier from $-b$ to $b$ is
situated ($b<a$).
 \begin{eqnarray}
\label{eq2.7}
 V(x)=\left\{
\begin{array}{ll}
\infty \hspace{2cm} & \mbox{if }\;|x|\ge a ,
\\0
\hspace{2cm} & \mbox{if } \;b\le| x |\le a,\\ V \hspace{2cm}
& \mbox{if }0\le |x|\le b \end{array} \right. \end{eqnarray}

We consider a wave packet which is, initially, concentrated on
one side of the barrier only, and in the course of time moves to
the other side of the
well\footnote{ Note that in the causal interpretation a particle
described by a
stationary wave function does not move,
and hence no tunneling occurs.}.\\

To obtain such a wave packet, we calculate, according to standard
procedure (see Appendix) the energy eigenvalues and eigenfunctions
for this double potential well. A linear combination of the lowest
even energy eigenfunction ($f_e(x,t)$) and the lowest odd energy
eigenfunction ($f_o(x,t)$) is the wave function considered in this
paper:
\begin{eqnarray}
\label{eq2.8} \Psi^{(0)}(x,t)=\frac{1}{\sqrt{2}}
f_e(x,t)+\frac{1}{\sqrt{2}} f_o(x,t)
\end{eqnarray}
The choice of these constants assures a high probability of
finding the electron at $t=0$ between $-a$ and $-b$.

Eqn. (\ref{eq2.8}) provides a wave packet with (approximately)
maximum extinction at one side. However,  this extinction is not
complete and can never be complete. We could enhance this
extinction by using a linear combination of more than two
eigenfunctions. But this might introduce recurrent
trajectories, which we want to avoid.\\

 Our wave packet shows periodic behavior, and transmission
from one side to the other takes place in half a period. This
 period time is inversely proportional to
$E_o^{(0)}-E_e^{(0)}$, where $E_o^{(0)}$ and $E_e^{(0)}$ are
proportional to the odd and even energy eigenvalues respectively.

From the wave packet we can calculate trajectories. The particle
velocity in the causal interpretation is given by \cite{Bohm:52,
Holland:93}:\begin{eqnarray} \label{eq2.3}
\dot{x}=\frac{\frac{\hbar}{m} Im(\Psi^*(x,t)
\frac{\partial}{\partial x}\Psi(x,t)}{\left | \Psi(x,t) \right
|^2}
\end{eqnarray}
where the right-hand side denotes the probability current density
over the probability density function. Analytical expressions of
the velocity ($\dot{x}$) can be obtained by using the expressions
given in the Appendix.

The trajectories are obtained as the solutions  to equation
(\ref{eq2.3}), subject to the specification of the initial
condition $x_0$. An ensemble of possible trajectories associated
with the same wave is generated by varying $x_0$. Hence, to obtain
the trajectories $x(t)$ the differential equation (Eqn.
\ref{eq2.3}) should be solved after a starting value ($x_0$) is
chosen. We solved this differential equation numerically
\cite{num}. The trajectories are given in Figs. \ref{fig2.2} and
\ref{fig2.3}. In the last figure trajectories are weighted by the
initial particle density $\rho(x,0)=|\Psi(x,0)|^2$.

\subsection{A classification of the trajectories}

\begin{figure}
\begin{center}
\epsfxsize=9cm
\epsfbox{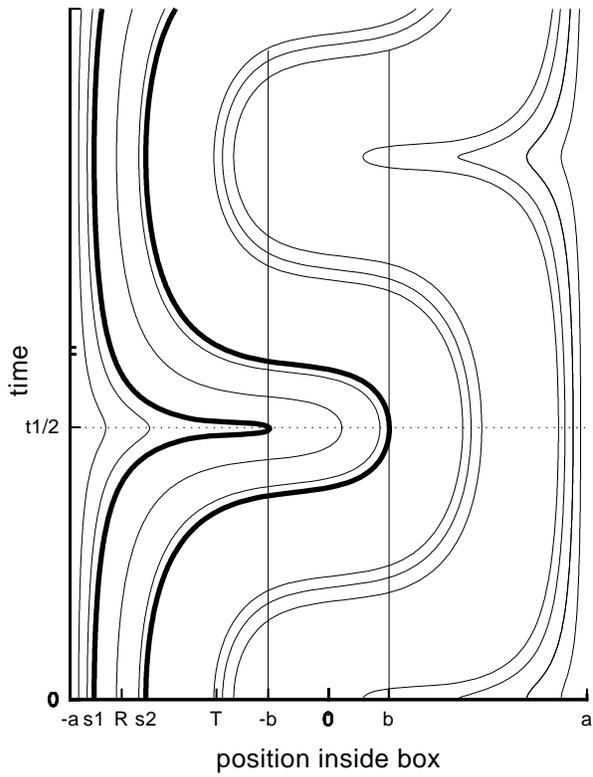}
\caption{\label{fig2.2} Bohm trajectories to show the different
starting point
$x_0$ possibilities at $t=0$. Left of $s_2$ and right of $b$
trajectories do not leave their own area. R=reflection area,
starting positions between $s_1$ and $s_2$. T=transmission area
starting positions between $s_2$ and $-b$.}
\end{center}
\end{figure}

Let us first note some salient aspects of the form of the possible
trajectories. Fig. \ref{fig2.2} shows various trajectories, that
are generated by varying the choice of an initial  position
($x_0$). One clearly sees their periodic behavior, and that the
trajectories do not cross. Note that at any given time, all
trajectories move in the same direction, i.e., to the right during
the first half period, and towards the left during the second
half.  The figure shows that at very left and at the very right
there are trajectories, which remain on the same side of the
barrier. This occurs for all trajectories with a starting point
left of  point $s_1$ and to the right of $b$. The area indicated
with R (returners) contains trajectories, which reach inside the
barrier but do not pass it, they return to their original area.
Their starting positions at $t=0$ is between $s_1$ and $s_2$.
These trajectories correspond to reflection. The trajectories
starting in area T (travellers) are trajectories, which pass the
barrier and end up, after half a period, at the other side of the
barrier. At $t=0$ the starting point of the travellers is between
$s_2$ and $-b$. This behavior corresponds to transmission. The
trajectories between $-b$ and $b$ are already inside and between
$b$ and $a$ already past the barrier at
$t=0$.\\

\begin{figure}
\begin{center}
\epsfxsize=11cm
\epsfbox{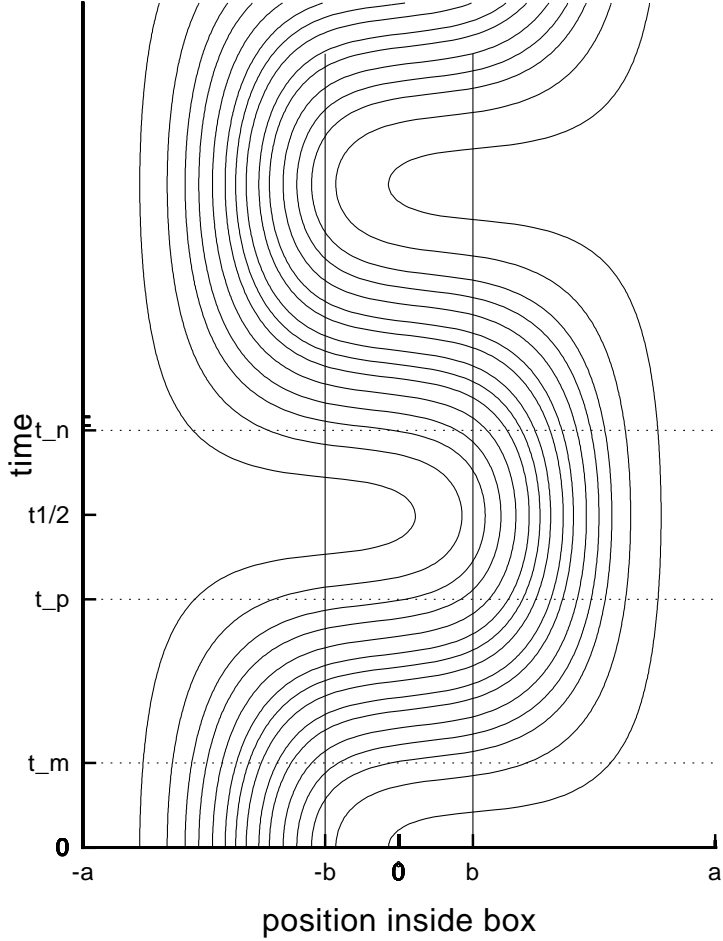}
\caption{\label{fig2.3} Bohm trajectories for a one dimensional
double potential
well.
We represent the probability density
function $t=0$ by N points. These points are spaced with equal
probability density intervals.
$\int_{-a}^{x_N}\,dx\left|\Psi(x,0)\right|^2=\frac{N-\frac{1}{2
}}{N_{tot}}$.
$x_N$ is the position of $x_0$ dependent on $N$. $N_{tot}$ is
the total number of points. The value of $-\frac{1}{2}$ is
arbitrarily chosen. Any other value (between 0 and 1) gives an
equally valid description. $t_p$ is the time the trajectory,
starting at $t=0$ in $s_2$, passes at $-b$.
$t_m=t_{\frac{1}{2}}-t_p$ and $t_n=t_{\frac{1}{2}}+t_m$. In this
example $N=15$.}
\end{center}
\end{figure}

In Fig.~\ref{fig2.3} the probability density $|\Psi(x,0)|^2$
according to the wave function (\ref{eq2.8}) is taken into account
(the meaning of the symbols $a$, $b$, $s_1$, $s_2$ and
$t_{\frac{1}{2}}$ are given in Figs. \ref{fig2.2} and
\ref{fig2.3}). Along the time axis we marked some special times,
$t_p$ $t_m$ and $t_n$. Time $t_p$ is the instant at which the
trajectory, starting at $t=0$ in $s_2$ passes at $-b$. This
trajectory marks the bifurcation between the reflection and
transmission area. Time $t_m$ indicates the time the trajectory
starting at $t=0$ in $-b$ arrives at $b$. This is the time needed
to deliver the trajectories, which at $t=0$ already are inside the
barrier, to the right hand side of the barrier. Although time
$t_n$ lies outside half a period of time, we marked this time
because it gives information about reflection times. All
trajectories starting at $t=0$ between $s_1$ and $s_2$ are at
$t=t_{\frac{1}{2}}$ inside the barrier. Between
$t=t_{\frac{1}{2}}$ and $t=t_n$ they leave the barrier and hence,
reflected particles are inside the barrier between time $t_p$ and
time $t_n$. \\

\subsection{Definitions of transmission and reflection
coefficients}

In view of the above classification, the most straightforward way
to define transmission and reflection seems to reserve the term
transmission for travellers (T) and reflection for returners (R).
Indeed, only
 travellers are actively involved in the transmission process.
A probability for transmission, the
transmission coefficient $\left| T \right|^2$, can then be
defined as:
\begin{eqnarray}
\label{eq2.10}
\left| T \right|^2=\int^{-b}_{s_2}\,dx \left| \Psi(x,0)
\right|^2 \end{eqnarray}
The reflection coefficient can be defined likewise:
\begin{equation} \label{eq2.11}
\left| R\right|^2=\int^{s_2}_{s_1}\,dx \left| \Psi(x,0)\right|^2
\end{equation}
Note that this reflection coefficient is in fact related to 
$\left| R^{'}\right|^2$ in relation (\ref{eq2.1a}). The
transmission and reflection coefficients (\ref{eq2.10} and \ref{eq2.11})
do not add up to unity:
$\left| T \right|^2+\left| R \right|^2\not= 1$.
Of course, this is due to the fact that we did not include 
$\left| R^{''}\right|^2$ and that there is a finite
probability that the trajectories are already inside or past
the barrier.
These last possibilities
are usually excluded in discussions of tunneling in
one-dimensional scattering processes \cite{Leavens:90}.

\subsection{Definitions of dwell time, transmission and
reflection times}

Average dwell time is the average time that particles spend
inside the barrier region:
\begin{eqnarray}
\label{eq2.12a}
\ave{t_d}=\int_0^{\infty} dt \int_{-b}^b \,dx \left | \Psi(x,t)
\right|^2
\end{eqnarray}
The probability to encounter a particle inside the barrier
($\int_{-b}^b \,dx \left | \Psi(x,t) \right|^2$) is for the wave
function given in Eqn. (\ref{eq2.8}) independent of time and
hence, the average dwell time
 inside the barrier in half a period is the probability to
encounter
a particle inside the barrier multiplied by  half a period:
\begin{eqnarray}
\label{eq2.12}
\ave{t_d}=t_\frac{1}{2} \int_{-b}^b \,dx \left | \Psi(x,t)
\right|^2
\end{eqnarray}
\\

To define average transmission and reflection times, we have to go
back to the trajectories because trajectories provide information
about the position ($x$) at each instant of time ($t$), which,
under the assumption that each trajectory passes $x$ only once,
can be inverted to give the function $t(x)$.  It is convenient to
express  the transmission and reflection times in terms of arrival
time distributions. In particular,  let the instant at which a
trajectory, starting in $x_0$ at $t=0$, arrives at $x_1$, be
denoted as $t(x_0;x_1)$. Averaging over the probability density
$|\Psi(x_0,0)|^2$ that a particle starts at
 $x_0$,
we obtain the
arrival time distributions ($\Pi(t)$):
\begin{eqnarray}
\label{eq2.13a}
\Pi_{x_1}(t)=\frac{\int\,dx_0|\Psi(x_0,0)|^2\delta(t-t(x_0;x_1))}
{
\int
\,dx_0|\Psi(x_0,0)|^2}
\end{eqnarray}
where the integration limits should be chosen in such a way as to
fulfill the assumption that each trajectory that passes $x_1$ does so only 
once in half a period.

 Hence, the arrival time distribution of the to-be-transmitted
particles at the exit of the
barrier $(b)$ is:
\begin{eqnarray}
\label{eq2.13}
\Pi_b(t)=\frac{\int ^{-b}_{s_2}
\,dx_0|\Psi(x_0,0)|^2\delta(t-t(x_0;b))}{ \int ^{-b}_{s_2}
\,dx_0|\Psi(x_0,0)|^2}
\end{eqnarray}
The denominator is the transmission coefficient (see formula
\ref{eq2.10}). The starting points $x_0$ are within the
transmission area, between $-b$ to $s_2$.\\
Similarly, the arrival time distribution of the to-be-transmitted
particles
at the entrance of the barrier $(-b)$ is:
\begin{eqnarray}
\label{eq2.14}
\Pi_{-b}(t)=\frac{\int^{-b}_{s_2}
\,dx_0|\Psi(x_0,0)|^2\delta(t-t(x_0;-b))}{ \int ^{-b}_{s_2}
\,dx_0|\Psi(x_0,0)|^2}
\end{eqnarray}

As long as the trajectories cross a particular point ($x_1$)
only
once,
Leavens \cite{Leavens:93} showed that:
\begin{eqnarray}
\label{eq2.15}
\int  \,dx_0|\Psi(x_0,0)|^2\delta(t(x_1)-t(x_1;x))=j(x_1,t(x_1))
\end{eqnarray}
which gives us the arrival time distributions in term of
probability current densities.
Hence, the average arrival time at point $x_1$
($\ave{t_a(x_1)}$)
under
the same conditions, is:
\begin{eqnarray}
\label{eq2.16}
\ave{t_a(x_1)}=\frac{\int dt \, t j(x_1,t)}{\int dt \, j(x_1,t)}
\end{eqnarray}

In our case, the probability current density in the double potential well
is
unidirectional for half a period and hence, we can determine the
average arrival time at the entrance and the exit of the
barrier.
Taking the time boundaries from Fig. \ref{fig2.3} in
consideration the average arrival time at the
entrance of the barrier is:
\begin{eqnarray}
\label{eq2.17}
\ave{t_a(-b)}=\frac{\int^{t_p}_0 dt \, t j(-b,t)}{\int^{t_p}_0
dt \, j(-b,t)}
\end{eqnarray}
and at the exit:
\begin{eqnarray}
\label{eq2.18}
\ave{t_a(b)}=\frac{\int^{t_{\frac{1}{2}}}_{t_m} dt \, t j(b,t)}
{\int^{t_{\frac{1}{2}}}_{t_m} dt \, j(b,t)}
\end{eqnarray}
and hence, the average transmission time, the average arrival
time at the exit of
the barrier minus the average arrival time at the entrance of
the
barrier, reads:
\begin{eqnarray}
\label{eq2.19}
\ave{t_t}=\ave{t_a(b)}-
\ave{t_a(-b)}=\frac{\int^{t_{\frac{1}{2}}}_{t_m} dt \, t j(b,t)}
{\int^{t_{\frac{1}{2}}}_{t_m} dt \, j(b,t)}-\frac{\int^{t_p}_0
dt \, t j(-b,t)}{\int^{t_p}_0 dt \, j(-b,t)}
\end{eqnarray}

To determine reflection times, we have to extend the observation
time to $t_n$ (see Fig. \ref{fig2.3}). To-be reflected particles
enter the barrier between time $t_p$ and $t_{\frac{1}{2}}$, they
pass the barrier again on their way back between time
$t_{\frac{1}{2}}$ and $t_n$. Hence, the average reflection time
is:
\begin{eqnarray}
\label{eq2.20}
\ave{t_{r^{'}}}=\frac{\int_{t_{\frac{1}{2}}}^{t_n}dt \, t j(-
b,t)}{\int_{t_{\frac{1}{2}}}^{t_n} dt \, j(-b,t)}-
\frac{\int_{t_p}^{t_{\frac{1}{2}}} dt \, t
j(-b,t)}{\int_{t_p}^{t_{\frac{1}{2}}} dt \, j(-b,t)}
\end{eqnarray}
In this definition we use time $t_n$, which does not fall inside
the range defined as half a period (from $t=0$ to
$t_{\frac{1}{2}})$. This might seem in contradiction to Eqn
(\ref{eq2.12}), the definition of average dwell time, which uses
$t_{\frac{1}{2}}$ as its upper bound limit. However, the time
taken by the reflecting particles on their way back to their
original place ($t_{\frac{1}{2}}$ to $t_n$) is equal to the time
taken by particles already inside the barrier at $t=0$, to leave
the barrier. The last mentioned time was not accounted for as
belonging to average transmission time. Hence, the integration
limits of the addition of average transmission and reflection
times and the total average dwell times are consistent.

\section{\label{sec1.5}Transmission time in terms of
probability density of
the wave function}

In the previous section, the definitions of the average
transmission and reflection times, using the average arrival time
distribution were based on trajectories from the causal
interpretation of quantum mechanics. However, the final
expressions (Eqns. \ref{eq2.19} and \ref{eq2.20}) do no longer
depend on the trajectories but on probability current densities.
This suggests the possibility to define average transmission (and
reflection times) without explicit use of the trajectories. In
this section, we will show, that average transmission times can,
indeed be obtained without explicit reference to trajectories.
Instead we only rely on the non-crossing property of the causal
interpretation and the unidirectionality between $t=0$ and
$t=t_{\frac{1}{2}}$ of the current density flow and the
periodicity of the wave functions. Average reflection times will
not be considered, because we are interested in the transmission
process. The question whether, the fact that the trajectories are
not needed explicitly in order to determine the transmission times
implies that the causal interpretation is superfluous for this
purpose is left for the next section.\\

\begin{figure}
\epsfxsize=10cm
\epsfbox{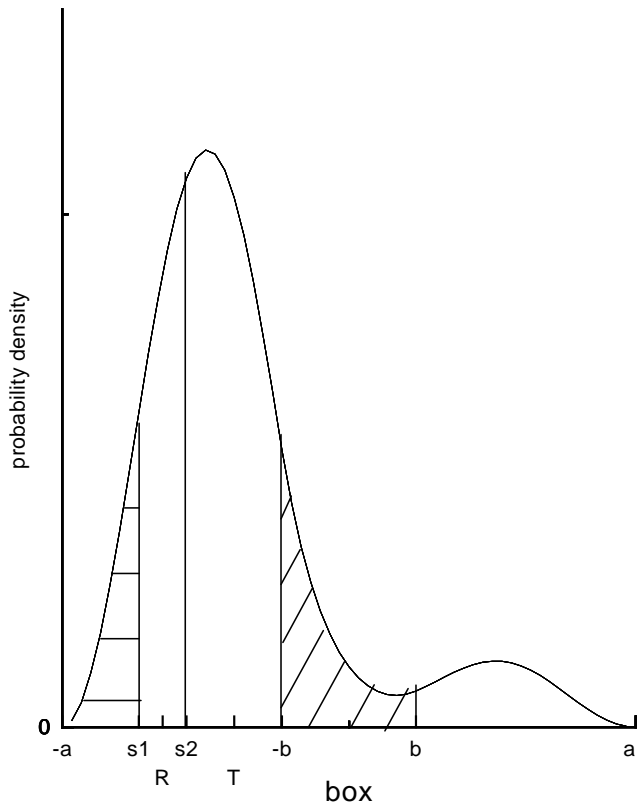}
\caption{\label{fig2.4} Probability density at $t=0$. The
transmission area
($T$) and reflection area ($R$, between $s_1$ and $s_2$) are
indicated. Their destinations are given in Fig. \ref{fig2.5}.
$-a$ and $a$
are box boundaries and $-b$ and $b$ the barrier boundaries.}
\end{figure}

\begin{figure}
\epsfxsize=10cm
\epsfbox{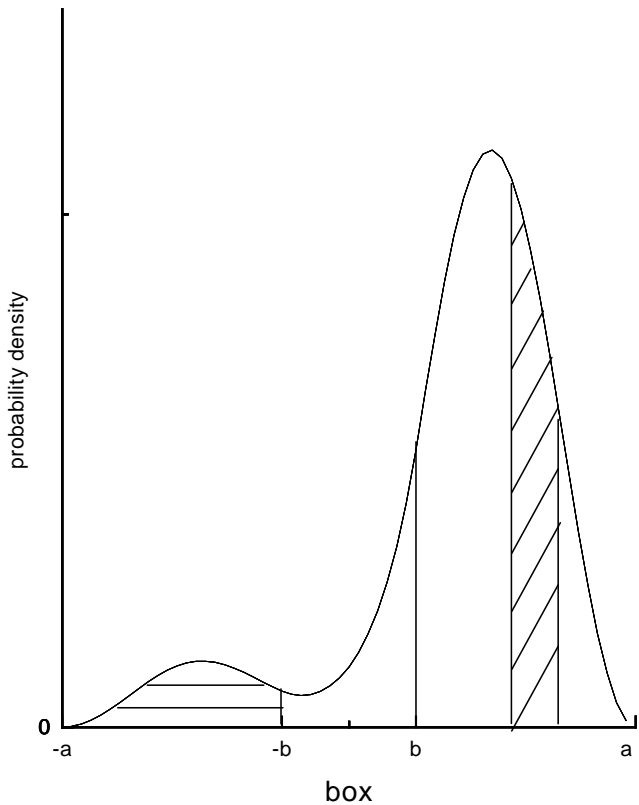}
\caption{\label{fig2.5} Probability density at
$t=t_{\frac{1}{2}}$. Compare to
Fig. \ref{fig2.4}: $T$ is past the barrier and $R$ inside the
barrier.}
\end{figure}

In order to show how average transmission times for a periodic
wave function can be determined from the probability density, we
refer to  Figs. \ref{fig2.4} and \ref{fig2.5} \footnote{Actually,
we chose the constants (see Eqn. (\ref{eq2.8})) in such a way that
the probability to encounter particles at $t=0$ at the right-hand
side of the barrier is large enough to be visible in graphical
presentations. }. Here we have partitioned the interior of the
double well in 5 areas, as shown in the figures.  In Fig.
\ref{fig2.4} the probability densities at $t=0$ and in Fig.
\ref{fig2.5} at $t=t_{\frac{1}{2}}$ are given. One can see that
the probability density goes from left to right during half a
period. The probability densities in Figs. \ref{fig2.4} and
\ref{fig2.5} are mirror images. We will
exploit this symmetry in our calculations.\\

The 5 areas indicated in Figs. \ref{fig2.4} and \ref{fig2.5} have
the
following meaning:\begin{itemize}
\item The probability in area at the utmost left-hand side,
between $-a$ and $s_1$,
remains
in its own domain.  $s_1$ can be found by the condition that the
probability between $-a$ and $s_1$ should be equal to the
probability between $-a$ and $-b$ in Fig. \ref{fig2.5}. Because
of symmetry the probability density between $-a$ and $s_1$ is
equal to the probability density between $b$ and $a$ at $t=0$
(Fig. \ref{fig2.4}). Hence,
\begin{eqnarray}
\label{eq2.29}
\int^{s_1}_{-a}\,dx \left| \Psi(x,0) \right|^2=\int^{a}_{b}\,dx
\left| \Psi(x,0) \right|^2
\end{eqnarray}

\item The area which is indicated with an $R$ in Fig.
\ref{fig2.4} is
the
reflection area (between $s_1$ and $s_2$). In Fig. \ref{fig2.5}
the probability within that area has moved inside the barrier.
The probability density inside the barrier is constant at all
times and hence is the same as the
probability inside the barrier at $t=0$:
\begin{eqnarray}
\label{eq2.30}
\int^{s_2}_{s_1}\,dx \left| \Psi(x,0) \right|^2=\int^{b}_{-b}\,dx
\left| \Psi(x,0) \right|^2
\end{eqnarray}\\

\item We indicated a $T$ in Fig. \ref{fig2.4} for the
transmission area (between $s_2$
and
$-b$). The probability density from this area arrives at
$t=t_{\frac{1}{2}}$, Fig. \ref{fig2.5}, at the opposite side of
the barrier. Because the probability density of all other areas
are known, this probability density can be determined by:
\begin{eqnarray}
\label{eq2.31}
\int^{-b}_{s_2}\,dx \left| \Psi(x,0)
\right|^2=1-2 \int^{a}_{-b}\,dx \left| \Psi(x,0)
\right|^2
\end{eqnarray}\end{itemize}

To determine the average transmission time, we need the
average arrival time of the to-be transmitted particles at the
entrance of the barrier and at
the exit of the barrier. We use the assumption that the flux is
unidirectional between $t=0$ and $t=t_{\frac{1}{2}}$.

\begin{itemize}
\item At the entrance of the barrier the transmission flux starts
at $t=0$ and should be stopped at the time, when the left-hand
side of the barrier is emptied of all travelling probability. The
probability left behind at the left-hand side of the barrier at
$t=t_p$ is equal to the probability between $-a$ and $s_2$ at
$t=0$. Hence, this time, $t_p$, can be found implicitly by:
\begin{equation}
\label{eq2.32}
\int^{-b}_{-a}\,dx \left| \Psi(x,t_p)
\right|^2=\int^{s_2}_{-a}\,dx
\left| \Psi(x,0) \right|^2
\end{equation}

\item At the exit of the barrier, we must wait until all the
probability initially inside the barrier has passed before the
travelling part arrives. This happens between time $t=0$ and
$t=t_m$. After half a period the travelling part is inside the
right-hand side well. Between time $t_m$ to $t_{\frac{1}{2}}$ the
 transmission part passes at
the exit of the barrier.
$t_m$ starts when the right hand side of the barrier contains the
probability of the right-hand side and the probability under the
barrier:
\begin{eqnarray}
\label{eq2.33}
\int^{a}_{b}\,dx \left| \Psi(x,t_m) \right|^2=\int^{a}_{-b}\,dx
\left| \Psi(x,0) \right|^2
\end{eqnarray}

The right-hand side terms of Eqns. (\ref{eq2.32}) and
(\ref{eq2.33}) are equal. The half period evolution
($t_{\frac{1}{2}}$) is $\frac{\pi}{E_1-E_0}$ (see the Appendix for
symbols) \footnote{Using the probability current density the times
$t_p$ and $t_m$ can also be found by:
\begin{eqnarray} \label{tomb} \left| T \right|^2= \int^{t_p}_0 dt
\, j(-b,t) =\int^{t_{\frac{1}{2}}}_{t_m} dt \, j(b,t)
\end{eqnarray}
}. \end{itemize}

Now, Eqn. (\ref{eq2.19}) can be filled in, which gives us the
average transmission time of the double
potential well, without using the
trajectories.\\

\section{Discussion and conclusions}
 In this section, we discuss the question whether or not the
causal interpretation provides an unambiguous way to define
average transmission times in a double potential well. Secondly,
we discuss the necessity to adopt the causal interpretation to
define this transmission time. We also discuss the
experimental accessibility of transmission times.

In the causal interpretation of quantum mechanics, the
position of an individual particle travelling along a particular
trajectory is determined at each instant of time. The trajectories
show the possible behavior of particles inside the barrier and
hence, the causal interpretation creates the possibility (at least
numerically) to discriminate between trajectories inside the
barrier, whose fate is transmission and whose fate is reflection.
Although definitions for transmission and reflection coefficients
are not common practice in a double potential well, the
definitions, given in Eqns. (\ref{eq2.10}) and (\ref{eq2.11})
provide a useful tool to differentiate between these two
possibilities. The wave functions
were made out of two eigenfunctions and hence the trajectories
show that the current density flow, between $t=0$ and
$t=t_{\frac{1}{2}}$, is unidirectional. Hence, a straightforward 
way to define transmission is provided. We took the lowest energy level 
eigenfunctions but any pair of eigenfunctions would give an equally 
straightforward way to define transmission.
However, the addition of more eigenfunctions to create wavefunctions would cause
recurrent trajectories. In case of recurrent trajectories the labeling of transmitted 
and reflected trajectories is not straightforward. \\

Similar to the more well-
known case of scattering processes, the transmission times can be
conveniently expressed in terms of the arrival time distributions.
Average arrival times at the entrance and exit of the barrier are
used to express the transmission time. Hence, the average time
spent inside the barrier by eventually transmitted particles can
(for the above mentioned wave functions) unambiguously be defined
and hence the answer to the question whether the causal
interpretation can give an unambiguous definition of average
transmission time in a
double potential well is "yes".\\

Next, let us discuss the question  whether or not the causal
interpretation of quantum mechanics is needed to define
transmission time.  Transmission times for a double potential well
are defined in the causal interpretation of quantum mechanics.
However, the causal interpretation also implies a different world
view than the standard, orthodox interpretation. The trajectories
describe the way particles move. The initial position of the
particle, although unknown to us, fixes its future path completely
and hence, in contrast to the orthodox interpretation, the causal
interpretation theory is a
deterministic theory. \\

In section \ref{sec1.5} we showed how the definition of
transmission times
could be obtained, without trajectories, from the probability
density of the wave function. This may suggest that
the causal interpretation is superfluous for the determination of
transmission times. However, transmission times are obtained
under the assumption that
 the flow of probability  density
follows the non-crossing property of the trajectories of causal
interpretation. Outside
the causal interpretation, the justification for this assumption
is not clear.
Hence, our discussion of the
definition of average transmission time in a double potential is
dependent on this aspect of the causal interpretation of quantum
mechanics for
its justification.  \\

Finally, we pose the
question of the experimental accessibility of average
transmission times.
For a double potential well model, a definition for the average
transmission time is offered in this paper but the question
whether
an experimental set-up to measure average transmission times can
be devised is open and hence, the question remains whether the
average transmission time for a double potential well can be
verified by experiments. The usefulness of the definition of
average transmission times would be greatly enhanced if average
transmission times can be
experimentally measured.

\section*{Acknowledgements}
I have benefited greatly from discussions with Jos Uffink and
Frank Vergeldt was of great help in performing programming
acrobatics.

\section*{Appendix}
\forget{
\subsection{\label{Ap2.A}Calculation of energy eigenvalues and
construction of the wave functions}} We take atomic units, i.e.\
we put $m=1$ and $\hbar$=1. Taking the boundary conditions in
consideration gives us even (symmetric) and odd (asymmetric)
solutions. For the even solutions we find:
\begin{eqnarray}
\label{eq2.22} f_e(x,t)= N_e ie^{-i E_e t}\left\{
\begin{array}{ll}
 \sin(k_e(x+a))  \hspace{2cm}    &

\mbox{if }\;  -a \le x \le-b , \\
\sin(k_e(b-a)) \frac{\cosh(\alpha_e x)}{\cosh(\alpha_e b)}
 \hspace{2cm}    &
\mbox{if }\;
         -b \le x \le b,\\
- \sin(k_e(x-a)) \hspace{2cm}    &
 \mbox{if }\;
\hspace{0.5cm}b\le x\le a  \end{array}
\right.
\end{eqnarray}

Similarly, for the odd solutions, one
obtains:
\begin{eqnarray}
\label{eq2.24} f_o(x,t)=N_o i e^{-i E_o t}\left\{
\begin{array}{ll}
 \sin(k_o(x+a))  \hspace{2cm}    &
\mbox{if }\;  -a \le x \le-b , \\
\sin(k_o(b-a)) \frac{\sinh(\alpha_o x)}{\sinh(\alpha_o b)}
\hspace{2cm}    & \mbox{if }\;
         -b \le x \le b,\\
\sin(k_o(x-a)) \hspace{2cm} & \mbox{if }\; \hspace{0.5cm}b\le x\le
a  \end{array} \right.
\end{eqnarray}
where $a$, $b$ and $V$ are explained in Fig \ref{eq2.1}, $N_{e,o}$ are (complex)normalisation factors,
$k_{e,o}=\sqrt{2E_{e,o}}$, $\alpha_{e,o}=\sqrt{2(V-E_{e,o})}$, and $E_{e,o}<V$.
Further, for even,
 $E_e$ is the solution of the equation:
\begin{eqnarray}
\label{eq2.23}
\arctan(\frac{\sqrt{E_e}}{\sqrt{V-E_e}}\coth(b\sqrt{2(V-E_e)})=n
\pi-(a-b)\sqrt{2E_e}
\end{eqnarray}
and for odd, $E_o$ is determined by
\begin{eqnarray}
\label{eq2.25}
\arctan(\frac{\sqrt{E_o}}{\sqrt{V-E_o}}\tanh(b\sqrt{2(V-E_o)})=n
\pi-(a-b)\sqrt{2E_o}.
\end{eqnarray}
\forget{ The total wave function is given in Eqn. (\ref{eq2.8}).}

\section*{Figure captions}
\subsection*{Fig. 1}
The  double potential well. The total length of the box is 2$a$,
the potential at $a$ and $-a$ is infinite. The barrier is situated
from$-b$ to $b$ and has a constant height V.
\subsection*{Fig. 2}
Bohm trajectories  showing the different types of behaviour,
depending on  starting point $x_0$ at $t=0$. to the left of $s_2$,
and right of $b$, trajectories do not leave their own area.
R=reflection area, starting positions between $s_1$ and $s_2$.
T=transmission area starting positions between $s_2$ and $-b$.
\subsection*{Fig. 3}
Bohm trajectories for a one dimensional double potential well. The
probability density function is represented at $t=0$ by N points.
These points are spaced with equal probability density intervals.
$\int_{-a}^{x_N}\,dx\left|\Psi(x,0)\right|^2=\frac{N-\frac{1}{2
}}{N_{tot}}$. $x_N$ is the position of $x_0$ dependent on $N$.
$N_{tot}$ is the total number of points. The value of
$-\frac{1}{2}$ is arbitrarily chosen. Any other value (between 0
and 1) gives an equally valid description. $t_p$ is the time the
trajectory, starting at $t=0$ in $s_2$, passes at $-b$.
$t_m=t_{\frac{1}{2}}-t_p$ and $t_n=t_{\frac{1}{2}}+t_m$. In this
example $N=15$.
\subsection*{Fig. 4}
The probability density at $t=0$. The transmission area ($T$)
and reflection area ($R$, between $s_1$ and $s_2$) are indicated.
Their destinations are given in Fig. \ref{fig2.5}. $-a$ and $a$
are box boundaries and $-b$ and $b$ the barrier boundaries.
\subsection*{Fig. 5}
 The probability density at
$t=t_{\frac{1}{2}}$.  In comparison with Fig. \ref{fig2.4}: $T$ is
now past the barrier and $R$ inside the barrier.

\end{document}